\documentstyle[aps,prl,epsf]{revtex}

\tighten
\begin{document}
\draft
\twocolumn[\hsize\textwidth\columnwidth\hsize\csname
@twocolumnfalse\endcsname


\title{Localization Length in Anderson Insulator with Kondo Impurities}

\author{ S.  Kettemann}

\address{ I. Institut f. Theoretische Physik, Universit\" at Hamburg,
 Hamburg 20355,  Germany}

\author{ M. E.  Raikh}

\address{ Department of Physics, University of Utah, Salt Lake City, 
Utah 84112, USA }
\maketitle

\begin{abstract}
The localization length, $\xi$,  in a 2--dimensional  Anderson insulator
 depends on  the  electron spin scattering rate by
magnetic impurities, $\tau_s^{-1}$. 
For antiferromagnetic sign of the exchange, 
the time  $\tau_s$ 
is  {\em itself a function of $\xi$},  
 due to the Kondo correlations. We demonstrate that  
 the  unitary regime of localization
 is   impossible when the concentration
of magnetic impurities, $n_{\mbox{\tiny M}}$, is smaller than a 
 critical value, $n_c$. 
For $n_{\mbox{\tiny M}}>n_c$, the dependence of $\xi$ on the
dimensionless conductance, $g$, is {\em reentrant}, 
 crossing over to   unitary, and back to  
 orthogonal behavior   upon 
increasing  $g$. 
Sensitivity of  Kondo correlations to a  weak {\em parallel}
magnetic field results in  a giant parallel
magnetoresistance.
\end{abstract}
\vskip2pc] {\em Introduction}. Scaling theory of localization
\cite{abrahams} renders predictions for the dependence of the
localization length, $\xi$, on the dimensionless conductance
$g=k_{\mbox{\tiny F}}l$, where $k_{\mbox{\tiny F}}$ is the Fermi
momentum, and $l$ is the mean free path.  
In two
dimensions, in the absence of a magnetic field, $\xi (g)$ increases with
$g$ as $\exp($$\pi g \over 2$$)$. Spin--flip processes caused by
magnetic impurities affect the $\xi (g)$ growth.  The underlying
reason is the suppression of the destructive interference, which,
without magnetic impurities, facilitates localization.  The quantitative
measure of this suppression is the spin--flip scattering time,
$\tau_s$, defined as
\begin{equation} \label{spinflip}
\frac{1}{\tau_s}=n_{\mbox{\tiny M}} \sigma_{sf}  v_{\mbox{\tiny F}},
\end{equation}
where $n_{\mbox{\tiny M}}$ is the concentration of  magnetic
impurities,
$\sigma_{sf}$ is the scattering cross section 
for a single magnetic impurity, and $v_{\mbox{\tiny F}}$ is the 
Fermi velocity.
The growth of $\xi(g)$ is significantly accelerated by magnetic
impurities when 
the conductance 
is high enough, so that $\xi(g)$
exceeds the spin--flip length, $(D\tau_s)^{1/2}$ \cite{nagaoka,larkin}, 
where $D=v_{\mbox{\tiny F}}^2\tau/2$
is the diffusion coefficient, and $\tau$
is the  scattering time. 
For higher $g$--values $\xi(g)$ rapidly  crosses over to 
$\exp(\pi^2 g^2)$--behavior, corresponding to the unitary ensemble.
The suppression of the weak localization correction 
 due to magnetic impurities was first pointed out 
 in Ref. \onlinecite{lee}.
  The fact that the first non-vanishing term in the scaling 
function is  $\sim 1/g^2$, suggesting the orthogonal--unitary crossover,
 was established in Refs.\cite{wegner,hikami}. 

The exact form of the dependence $\ln\xi(g)$ in the presence of the
spin--flip scattering cannot be inferred from the scaling 
theory.
Still it is possible to utilize the approach 
of Lerner and Imry \cite{lerner}, who considered 
the orthogonal--unitary crossover  with magnetic field,
and apply it to the case of the spin--flip scattering.
 This yields
\begin{equation} \label{crossover}
\ln\left(\frac{\xi}{l}\right) =\frac{1}{2}
\ln\left(\frac{\tau_s}{\tau}\right) 
+ \left[\pi g  - \ln\left(\frac{\tau_s}{\tau}\right)\right]^2,   
\end{equation} 
 Eq. (\ref{crossover}) emerges upon integration of the
scaling equation from the largest length, $\xi$, to the
smallest length, $l$. In course of integration,
the unitary form of the $\beta$-function must be used
within the interval $\left[\xi,(D\tau_s)^{1/2}\right]$
and the  orthogonal form within the interval
$\left[(D\tau_s)^{1/2}, l \right]$. 
A crossover equation similar to Eq. (\ref{crossover}) emerges
from  the calculation of the  
$\beta$--function following Ref.\cite{efetov}, but with $\tau_s$ kept
finite. 

In previous considerations it was implicit that  magnetic
impurities are free to flip their spins.  More precisely, the ability
to flip spins was assumed independent of the degree of localization of
the surrounding electrons.  On the other hand, for antiferromagnetic
sign of the exchange, it is evident that at zero temperature the
spin--flip scattering is completely suppressed due to the presence of
the Fermi sea.  Indeed, as the temperature is lowered, the screening
of spin of the magnetic impurity  by the Fermi sea electrons drives 
the impurity into the Kondo state, in which only spin--conserving 
scattering is possible\cite{tsvelik,andrei}.  
Thus, at zero temperature, 
the spin--flip scattering rate is expected to vanish, and  hence, no  
orthogonal--unitary crossover in  $\xi(g)$.

Our main point here is that the above 
scenario {\em does not} apply to the
Anderson insulator. The reason for this is the following.  
In the localized regime,
a single magnetic impurity "communicates" only with surrounding
electrons within a spatial domain  $\sim \xi$.  In other
words, the magnetic impurity in the {\em insulating environment} can
be viewed as a Kondo--box\cite{kroha} of  size $\sim \xi$.   
As a result, the ground state
of this impurity is determined by the ratio of two energies, namely
$T_{\mbox{\tiny K}}$ -- the Kondo temperature in the bulk, and
$\Delta_c$ -- the level spacing in the box.  In the case of the
Anderson insulator, we have $\Delta_c = \left(\nu \xi^2\right)^{-1}$, 
where $\nu$ is
the density of states\cite{remark}.  This leads us to the conclusion that the
spin--flip scattering rate at {\it zero temperature} can be obtained
from the finite--temperature expression\cite{tsvelik,andrei} with
temperature, $T$, replaced by $\Delta_c$. Therefore, the time, $\tau_s$,
which determines the localization length, $\xi$, through
Eq. (\ref{crossover}), is {\em itself} a function of $\xi$.  
With $\tau_s$ being $\xi$-dependent, Eq. (\ref{crossover}) 
becomes an {\em equation} for
the localization length.  As we will demonstrate below, this leads to
a nontrivial behavior of $\xi(g)$.  


{\em Localizaion length}. In order to find $\xi(g)$ from  
Eq. (\ref{crossover}),
the dependence   $\tau_s(\xi)$ should be  specified.
This can be done in two limiting cases 

\noindent ({\em i}) underdeveloped Kondo regime,
$T_{\mbox{\tiny K}} < \Delta_c$. In this limit the
leading logarithm 
approximation\cite{abrikosov65,suhl65} applies. Then, for the spin--flip
scattering cross section, we can use the 
 expression   derived with leading logarithm accuracy      \cite{zittartz,maple}
 and replace 
$T \rightarrow \Delta_c$. For the
impurity spin $S=1/2$ this yields
$\sigma_{sf}^{-1}=2k_{\mbox{\tiny F}}\left[1+\frac{4}{3 \pi^2}
\ln^2\left(\Delta_c/T_{\mbox{\tiny K}}\right)\right]$. 
It is convenient to rewrite the above expression in terms of 
the ratios $\tau_s/\tau$ and $\xi/l$ that enter the crossover
equation Eq. (\ref{crossover}). Using Eq. (\ref{spinflip}), we obtain
\begin{equation}
\label{ratio1}
\left.\frac{\tau_s}{\tau}\right\vert_{\Delta_c \gg  T_{\mbox{\tiny K}}}
\!\!\!=
\frac{4\pi n_e}{g n_{\mbox{\tiny  M}}}
\left\{\!1+\!\! \frac{4}{3\pi^2}
\ln^2\!\!\left[\left(\frac{\xi}{l}\right)^{-2}\!\!\left(
\frac{2\pi E{\mbox{\tiny F}}}{g^2 T_{\mbox{\tiny K}}}\right)\!\right]
\!\right\},
\end{equation}
where $n_e = k_{\mbox{\tiny F}}^2/2\pi$ is the concentration of 
electrons, and $E_{\mbox{\tiny F}}=
\hbar k_{\mbox{\tiny F}}v_{\mbox{\tiny F}}/2$ is the Fermi energy.

\noindent({\em ii}) fully developed Kondo regime, 
$T_{\mbox{\tiny K}}>\Delta_c$.
In this regime the spin of the magnetic impurity is screened, 
so that the spin--flip scattering  is suppressed. 
The dependence of $\tau_s$ on $\xi$ can be inferred from the following
reasoning. In a {\em homogeneous} system at temperature 
$T \ll T_{\mbox{\tiny K}}$ 
the full scattering cross section for 
an electron with energy $\sim T$
by a magnetic impurity
differs from the unitary value 
by a fraction 
$\sim \left(T/T_{\mbox{\tiny K}}\right)^2$.
This deviation
comes from two, comparable to each other,  
spin-conserving and  spin-flip contributions. 
In our case the temperature is zero, 
but the magnetic impurity is effectively located in the ``box'' 
of a size $\sim \xi$ with discrete
electron levels spaced in energy by 
$\Delta_c \ll T_{\mbox{\tiny K}}$.  
Using $\Delta_c$ instead of $T$ as the smallest energy scale, 
yields the 
following expression for $\tau_s$ in the fully developed Kondo regime
\begin{equation}
\label{ratio2}
\left.\frac{\tau_s}{\tau}\right\vert_{\Delta_c \ll  T_{\mbox{\tiny K}}}=
\frac{4\pi}{g}\left(\frac{n_e}{n_{\mbox{\tiny  M}}}\right)
\left(\frac{ T_{\mbox{\tiny K}}}{\Delta_c}\right)^2=
\frac{g^3n_e}{\pi n_{\mbox{\tiny  M}}}
\left(\frac{ \xi^2 T_{\mbox{\tiny K}}}
 {l^2 E_{\mbox{\tiny F}}}\right)^2,
\end{equation}
where the 
coefficient $4\pi$ in the first identity is chosen in
such a way that Eqs. (\ref{ratio1}) and (\ref{ratio2}) match at
$\Delta_c \approx  T_{\mbox{\tiny K}}$.
Replacement of $T$ by $\Delta_c$ in Eq. (\ref{ratio2}) warrants 
a more detailed discussion. Indeed, after being placed into the box, 
the ability of the magnetic impurity to flip 
the spin depends on the actual disorder realization 
within the  box  via the parity effect\cite{kroha}. 
In other words, the deviation from complete screening 
fluctuates randomly between zero and $\Delta_c/T_{\mbox{\tiny K}}$.
Finite $T$, on the other hand, causes the deviation from the complete 
screening by the amount $\sim T/T_{\mbox{\tiny K}}$.
The similarity between the situations
with finite $T$ and finite $\Delta_c$ is furthered 
by the observation that formation of the Kondo
state in the box with even number of electrons 
involves partial occupation of the first level 
above the Fermi level\cite{kroha}. The same happens when 
the temperature is of the order of $\Delta_c$.

The dependence $\tau_s(\xi)$ defined by Eqs. (\ref{ratio1}),
(\ref{ratio2}) is depicted schematically in Fig. 1a. The minimum 
corresponds to $\xi = \xi_{\mbox{\tiny K}}$, where the length 
$\xi_{\mbox{\tiny K}}$ is defined as
\begin{equation}
\label{xiK}
\xi_{\mbox{\tiny K}} = 
\left(\frac{ E_{\mbox{\tiny F}}}
{n_e T_{\mbox{\tiny K}}}\right)^{1/2}.
\end{equation} 
At $\xi = \xi_{\mbox{\tiny K}}$ we have 
 $\tau_s/\tau \approx 4\pi n_e /gn_{\mbox{\tiny  M}}$. 
Fig.~1a allows us to draw certain conclusions about the behavior of
the localization length as a function of the conductance. Indeed,
if the minimum lies above the bisector 
$\tau_s/2\tau = \left(\xi/l\right)^2$, then
the $\xi(g)$-- dependence remains ``orthogonal'', 
$\ln\left(\xi/l\right)=\pi g/2$, at any $g$. The corresponding 
condition can be rewritten as $n_{\mbox{\tiny M}} < n_c$, where
\begin{equation}
\label{critical}
n_c = \frac{n_e}{\pi}\left(\frac{T_{\mbox{\tiny K}}}
{E_{\mbox{\tiny F}}}\right)\ln \left(\frac{E_{\mbox{\tiny F}}}
{T_{\mbox{\tiny K}}}\right),
\end{equation}
is the critical concentration of magnetic impurities.
When $n_{\mbox{\tiny M}}$
exceeds $n_c$,
the line $\tau_s/2\tau = \left(\xi/l\right)^2$ intersects the 
$\tau_s/\tau$ curve
at two points, $\xi_1/l = \exp\left(\pi g_1/2\right)$ and 
$\xi_2/l = \exp \left(\pi g_2/2\right)$. Therefore, the 
orthogonal behavior of $\xi(g)$ holds within the domains
$g < g_1$ and $g > g_2$, as illustrated in Fig. 1b. 
With logarithmic accuracy the values $g_1$ and $g_2$ are given by
\begin{equation}
\label{g1}
g_1 = 
\frac{1}{\pi}\ln \left(\frac{2 \pi n_e}{n_{\mbox{\tiny M}}}\right),~~
g_2 = \frac{1}{\pi}
\ln\left[\left(\frac{\pi n_{\mbox{\tiny M}}}{n_e}\right)
\left(\frac{E_{\mbox{\tiny F}}}{T_{\mbox{\tiny K}}}\right)^2\right].
\end{equation}
In the domain $g_1 < g < g_2$ the dependence $\xi(g)$ takes
different forms within the intervals $g_1 < g < g_{\mbox{\tiny K}}$
and $g_{\mbox{\tiny K}} < g < g_2$, as shown in Fig. 1b.
The crossover conductance, $g_{\mbox{\tiny K}}$, is determined
from Eq. (\ref{crossover}) by substituting 
$\xi =\xi_{\mbox{\tiny K}}$ and using Eq. (\ref{ratio1}) for
$\tau_s/\tau$. This yields
\begin{equation}
\label{gK}
g_{\mbox{\tiny K}}=\frac{1}{\pi}
\left\{\ln \left(\frac{4\pi n_e}{n_{\mbox{\tiny M}}}\right)
+ 
\frac{1}{2^{1/2}}\ln^{1/2}\left(\frac{n_{\mbox{\tiny M}}
E_{\mbox{\tiny F}}}{ 2 n_e T_{\mbox{\tiny K}}}\right)
\right\}.
\end{equation}
In the interval  $g_1 < g < g_{\mbox{\tiny K}}$ the $\xi(g)$
dependence follows Eq. (\ref{crossover}) (see Fig. 1b). 
Further growth of $\xi$ with conductance is slow due 
to the suppression of the spin--flip
scattering in the fully developed Kondo regime. The analytical 
form of $\xi(g)$  in the interval $g_{\mbox{\tiny K}} < g < g_2$
can be found from Eq. (\ref{crossover}) 
upon substituting Eq. (\ref{ratio2}) for $\tau_s$ and solving
for $\xi$. We obtain
\begin{equation}
\label{slow}
\ln\left(\frac{\xi}{l}\right)= \frac{\pi g}{4} + \frac{1}{4}
\ln\left(\frac{\pi n_{\mbox{\tiny M}}E_{\mbox{\tiny F}}^2}
{n_e T_{\mbox{\tiny K}}^2}\right) =
\frac{\pi(g+g_2)}{4}.
\end{equation}
\begin{figure}
\centerline{
\epsfxsize=1.95in
\epsfbox{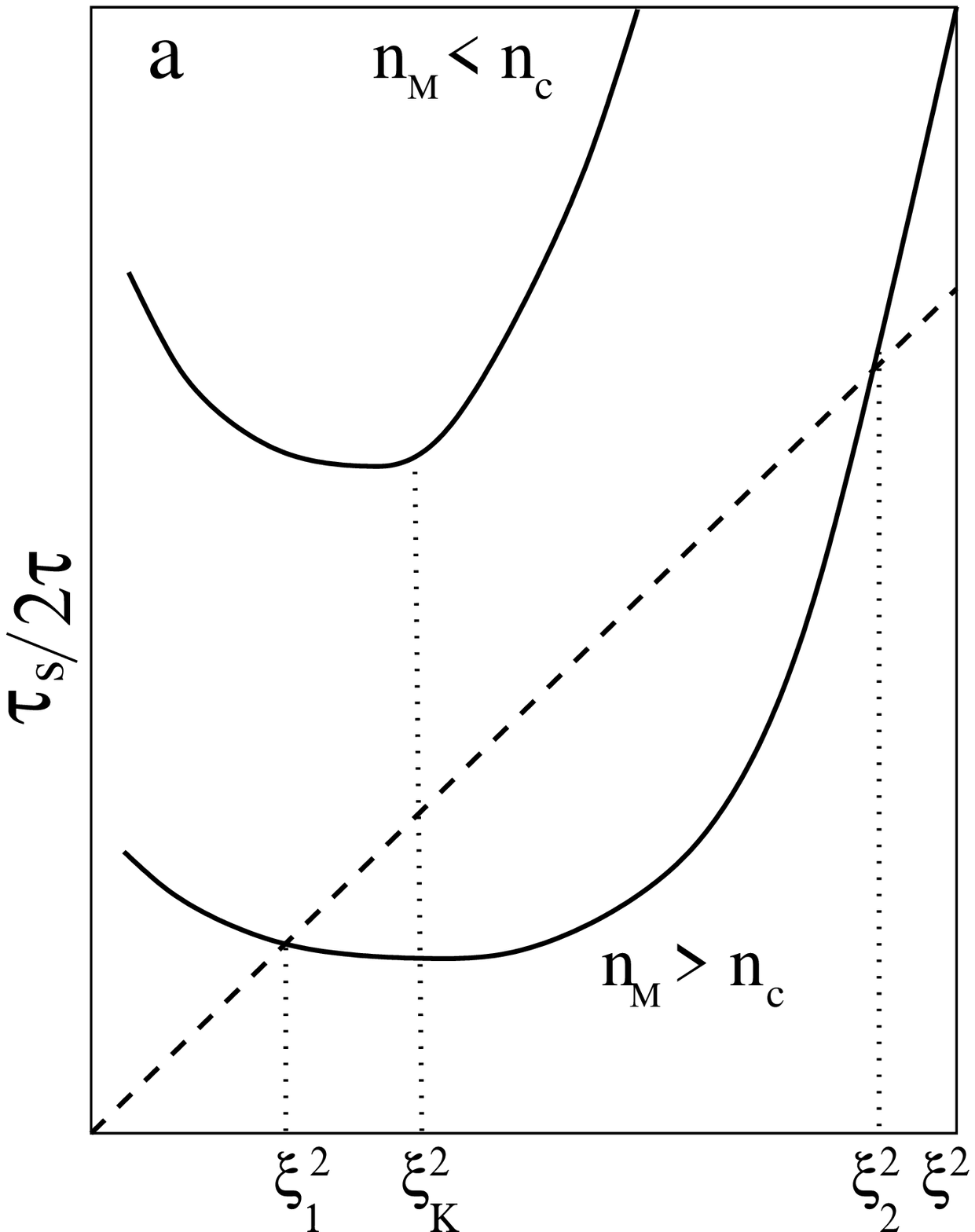}
\hspace{-.9cm}
\epsfxsize=1.78in
\epsfbox{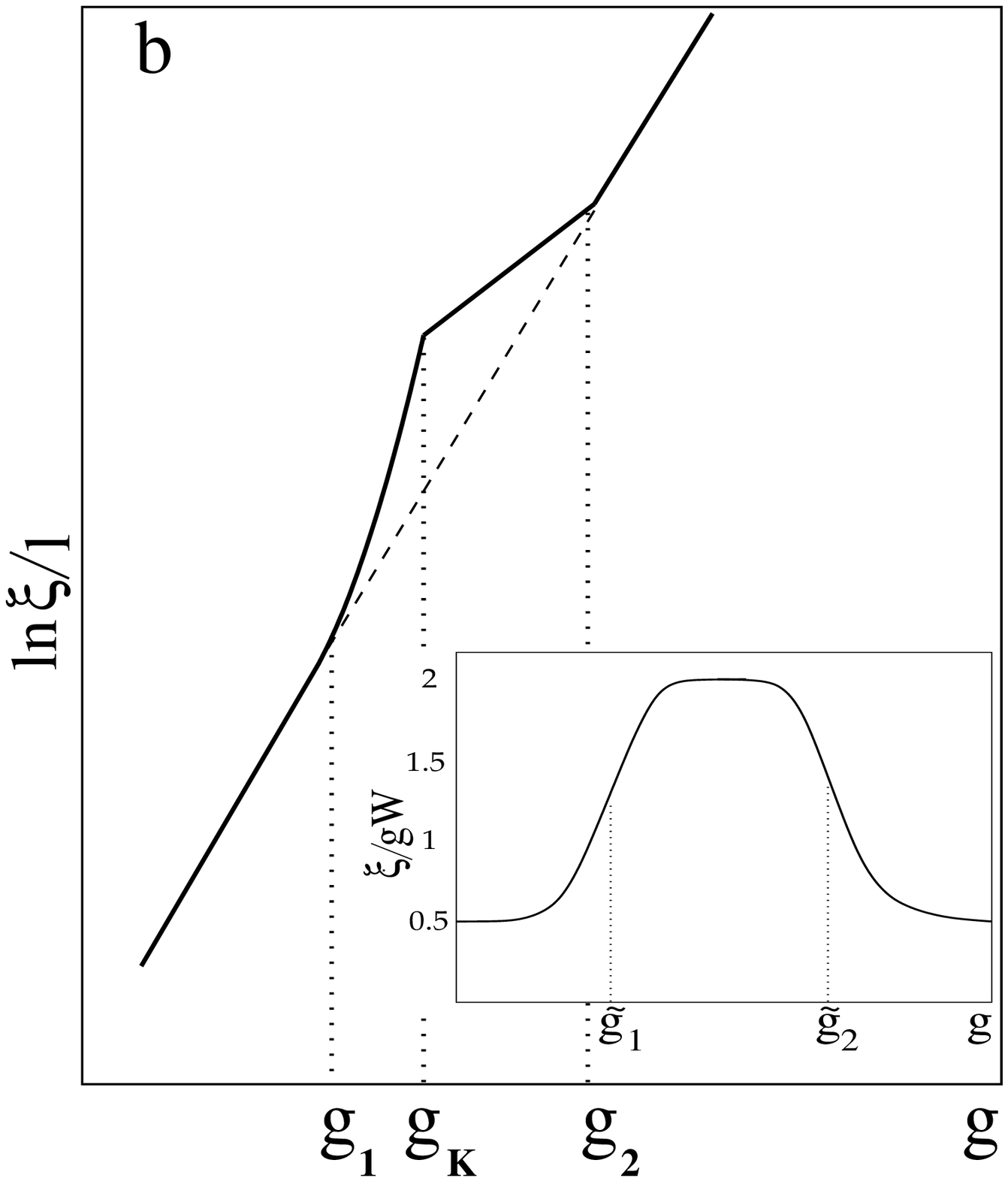}}
\protect\caption[sample]
{\sloppy{ (a) Dependence of the dimensionless spin scattering time
on the localization length (in the units of $l$) as defined by 
Eqs. (\ref{ratio1}), 
(\ref{ratio2}) is depicted schematically for different concentrations
of magnetic impurities, $n_{\mbox{\tiny M}}$; (b) localization
length, $\xi$, is plotted versus dimensionless conductance for 
$n_{\mbox{\tiny M}} > n_c$. For $g < g_1$ and $g > g_2$ the straight
line reflects the ``orthogonal'' behavior 
$\ln\left(\xi/l\right)= \pi g/2$. In the interval 
$g_1 < g < g_{\mbox{\tiny K}}$ localization length follows the
crossover formula Eq. (\ref{crossover}). Novel regime Eq. (\ref{slow})
due to the Kondo effect takes place within the interval
$g_{\mbox{\tiny K}} < g < g_2$. 
 The inset shows the   dimensionless localization length 
 in a quasi-one-dimensional geometry, 
 plotted schematically versus dimensionless
conductance for $n_{\mbox{\tiny M}} > \tilde n_c$. 
}}
\label{figone}
\end{figure}
Eq. (\ref{slow}) is our central result. It describes the intermediate 
behavior of the localization
length, when it is much smaller than the ``unitary'' value, but
much larger than the ``orthogonal'' value (see Fig. 1b). 
The slope, 
$\partial \ln \xi/\partial g = \pi/4$ in this regime
is $2$ times smaller than in the orthogonal  regime. The factor $2$ 
originates from the $\xi$-dependence
of the r.h.s. of Eq. (\ref{ratio2}) and reflects the underlying
physics, namely, that the ability of the Kondo impurity to flip the
spin is governed by the level spacing in the Kondo--box\cite{kroha}.

We  found that the spin--flip induced 
enhancement of the length $\xi$
occurs when the concentration
of magnetic impurities exceeds $n_c$ given by Eq. (\ref{critical}).
At $n_{\mbox{\tiny M}} = n_c$, the condition 
$\tau_s/2\tau =\left(\xi/l\right)^2$
is satisfied for $\xi = \xi_{\mbox{\tiny K}}$. For consistency of our
consideration
the number of magnetic impurities within
the area $\xi_{\mbox{\tiny K}}^2$ must be large. Indeed, from 
Eqs. (\ref{critical}), (\ref{xiK}) we find 
$n_c~\xi_{\mbox{\tiny K}}^2 \sim 
\ln\left(E_{\mbox{\tiny F}}/T_{\mbox{\tiny K}}\right) \gg 1$. 

{\em Quasi-1D geometry}. The above consideration can be easily 
extended to the 
quasi-one-dimensional geometry, i.e. 
a wire of 
a width $W$. The difference between the orthogonal
and unitary regimes  is less pronounced 
in this geometry. Namely, with increasing $g$
the localization length crosses over from $\xi(g) = gW/2$
to $\xi(g)= 2gW$. 
The crossover equation, analogous to Eq. (\ref{crossover}), 
is known\cite{kettemann1}.   
Analysis of this equation is very similar to the 2D
case. In this analysis the  expressions
(\ref{ratio1}), (\ref{ratio2})  for $\tau_s/\tau$ must be 
modified since for a wire
the level spacing in the
Kondo--box  
takes the
form $\Delta_c = \left(\nu \xi W \right)^{-1}$.  
We will outline the results of this analysis.
Presence of magnetic impurities affects $\xi(g)$
if the impurity concentration, $n_{\mbox{\tiny M}}$, exceeds 
$\tilde n_c \sim n_e \left(T_{\mbox{\tiny K}}/
E_{\mbox{\tiny F}}\right)$.
For $n_{\mbox{\tiny M}}> \tilde n_c$ 
the $\xi(g)$ dependence is depicted
schematically in  Fig. 1b, inset. The first crossover  at 
$\tilde g_1 = \left(n_{\mbox{\tiny M}}W^2\right)^{-1}$
 is a conventional
orthogonal--unitary crossover.
The second crossover  back to the orthogonal result occurs  
 at 
$\tilde g_2 =n_{\mbox{\tiny M}}\left(E_{\mbox{\tiny F}}/
n_e W T_{\mbox{\tiny K}}\right)^2$,
when the Kondo  regime is 
fully 
developed.

{\em The effect of RKKY exchange}. As we have established above,
 the localization length exceeds the
``orthogonal'' value only when 
$n_{\mbox{\tiny M}} > n_c$. 
The upper limit on $n_{\mbox{\tiny M}}$,
for which our consideration applies, 
comes from the indirect RKKY interaction, 
which leads to the strong correlation of
spins of  neighboring magnetic impurities (spin glass) and, thus,
precludes the formation of the Kondo states for individual impurities.
In order to incorporate the RKKY interaction, $E_{in}(r)$, into
the theory, we note that, even if {\em on average} this interaction
exceeds $T_{\mbox{\tiny K}}$, a certain fraction, $f$, of 
the impurities 
still remains in the Kondo regime. This fraction
is equal to
\begin{equation}
\label{fraction} 
f =\exp\left(-\pi n_{\mbox{\tiny M}} r_{\mbox{\tiny K}}^2\right),
\end{equation}
 where
$r_{\mbox{\tiny K}}$ is  determined from the condition
 $E_{in}(r_{\mbox{\tiny K}}) = T_{\mbox{\tiny K}}$. Then 
$f n_{\mbox{\tiny M}}$ is the concentration of the {\em isolated}
magnetic impurities, for which the indirect interaction with
{\em all} the  other impurities is smaller than $T_{\mbox{\tiny K}}$.
Thus, taking into account the RKKY exchange between the magnetic
impurities amounts to the replacement of $n_{\mbox{\tiny M}}$
in 
Eqs. (\ref{spinflip}), (\ref{ratio1}), (\ref{ratio2}), and (\ref{g1}--
\ref{slow})
by $fn_{\mbox{\tiny M}}
= n_{\mbox{\tiny M}}
\exp\left(-\pi n_{\mbox{\tiny M}}r_{\mbox{\tiny K}}^2\right)$.
The interval of $n_{\mbox{\tiny M}}$, where the above consideration
applies, is determined by the condition $f n_{\mbox{\tiny M}} = n_c$.
The latter product has a maximum, as a function of $n_{\mbox{\tiny M}}$,
  at $n_{\mbox{\tiny M}}=
\left(\pi r_{\mbox{\tiny K}}^2\right)^{-1}$. Thus, the interval of
validity exists if $\pi r_{\mbox{\tiny K}}^2 < n_c^{-1}$.
It is known\cite{zyuzin} that the disorder does not 
 affect dramatically  the decay law of the indirect exchange.
Therefore, the dependence $E_{in}(r)$ can be approximated as
$E_{in}(r) \approx E_{in}(0)/n_er^2$, where $E_{in}(0)$ is the
interaction at distances $\sim n_e^{-1/2}$. Using the expression
Eq. (\ref{critical}) for $n_c$, the condition of validity can
be rewritten in the form 
$E_{in}(0) < E_{\mbox{\tiny F}}/\ln\left(E_{\mbox{\tiny F}}
/T_{\mbox{\tiny K}}\right)$. On the other hand, 
since the RKKY exchange
integral 
also determines the
exponent in the Kondo temperature,
we can express 
$E_{in}(0)$ through $T_{\mbox{\tiny K}}$.
This yields the estimate $E_{in}(0) \sim E_{\mbox{\tiny F}}/
\ln^2\left(E_{\mbox{\tiny F}}/T_{\mbox{\tiny K}}\right)$.
Thus, 
parametrically, the interval of validity of
our theory exists, although the ``large'' parameter,
$\ln\left(E_{\mbox{\tiny F}}/T_{\mbox{\tiny K}}\right)$, is certainly
not reliable.    




{\em Implications}. Conventionally, the prime manifestation of the 
Kondo effect is
a drop of the Drude conductivity with decreasing temperature.
This behavior  reveals the temperature dependence of the
{\em full} scattering cross section from a Kondo impurity.  
Compared to this prominent effect,
other aspects of the Kondo physics  received much less 
attention in the literature. These more delicate
aspects were addressed in connection
with the pair breaking in superconductors\cite{zittartz},
weak localization corrections to the conductivity of metallic films
and wires\cite{bergmann2}, 
and, most recently, in connection with electron-electron interactions
mediated by Kondo impurities
\cite{pothier,kaminskii,anthore,goeppert1}.
In the case of the Anderson insulator, the  Kondo effect  
should manifest itself through the inelastic transport.
Indeed, the Mott law for the low--temperature resistance 
of the Anderson insulator
can be written as $\ln R \sim \left(\Delta_c/T\right)^{1/3}
= \left[\nu T \xi^2(g) \right]^{-1/3}$,
and applies at temperatures $T \ll \Delta_c$. As 
we demonstrated
above, Kondo impurities govern the dependence $\xi(g)$. 
This immediately suggests that the resistance of the Anderson 
insulator is 
exponentially sensitive  to a weak parallel
magnetic field, which otherwise has no effect on electronic states.
Indeed, for large enough $\xi$, 
so that $\Delta_c \lesssim  T_{\mbox{\tiny K}}$, a Zeeman splitting, 
$\Delta_{\mbox{\tiny Z}} \sim \Delta_c$, suppresses the Kondo 
correlations and, thus, causes a growth of $\xi$. 
In general, inelastic transport is a natural arena to test 
the scaling theory\cite{abrahams}. For example, 
a predicted doubling of $\xi$ in quasi-1D systems in a weak 
perpendicular 
field\cite{efetovlarkin} has been revealed
through the giant negative 
magnetoresistance in the localized regime\cite{gershenson}.
Correspondingly, we predict a giant negative magnetoresistance for
$g > g_{\mbox{\tiny K}}$ and a strong {\em positive} magnetoresistance
 for $g < g_{\mbox{\tiny K}}$. When discussing  inelastic transport, it
is important to remember that
our  key equation Eq. (\ref{ratio2}) 
implies
 that a single-particle 
localization length 
 has a meaning only for electron energy $\lesssim \Delta_c$. 
This, however, does not
restrict the applicability of the Mott law, since the typical
activation energy $\sim \left(\Delta_c T^2\right)^{1/3}$ is
much smaller than $\Delta_c$. 
Note, that the spin-glass state of magnetic impurities
in the Anderson insulator should be also sensitive to a
weak parallel magnetic field.
Since in the experiment, $n_{\mbox{\tiny M}}$
is never known precisely, there is a question
as to how to distinguish  the Kondo-state
of impurities (with spins completely screened)
from the glassy state (in which the spins are frozen).
In the  experimental paper \onlinecite{levy}, the
{\em mesoscopic} measurements of the asymmetric
(with respect to the reversal of magnetic field)
component of the resistance allowed the authors to
identify the spin-glass state. This asymmetric
component originates from the fact that the
time-reversal symmetry in the spin glass is broken.

{\em Conclusion}. Scaling theory of 
localization is a
scheme that accounts for {\em all} orders in the disorder strength.  The 
insightful observation \cite{nagaoka,wegner,hikami}
that interference
effects 
and the localization which they cause, 
 survive in the presence
of the spin--flip processes,
suggests that magnetic impurities can be 
included into the scaling theory through the crossover parameter 
$\left(\tau_s/\tau\right)\left(\xi/l\right)^{-2}$.
In the present paper the many--body physics is incorporated into this
scheme in a nonperturbative fashion,  i.e. with full account for
the interactions. This is achieved at the expense of the assumption
that the
 Coulomb 
interaction between the electrons is 
screened due to, say, a gate electrode. The only type of interactions
retained is the on--site Hubbard repulsion\cite{anderson}, which is
responsible for the magnetic properties of an impurity, and gives rise
to the Kondo physics in the presence of the Fermi sea.  Retaining only
the Hubbard repulsion, allows one to trace the interplay of the
interference and interaction effects on a nonperturbative level. Our
study suggests that this interplay results in the enhancement of
localization.

{\em Acknowledgments.}  The authors acknowledge the discussions with
I. L. Aleiner,  N. Birge, L. I. Glazman, and B. Kramer, and the hospitality of ICTP
 at Trieste.
S. K. acknowledges the discussions
 with A. Chudnovskiy, J. v. Delft, K. B. Efetov, I. V. Lerner,
 H. Kroha, and P. Mohanty.  M. R. acknowledges the hospitality of the
 I. Institut f\"ur Theoretische Physik at Universit\"at Hamburg.  This
 research was supported  
by EU TMR--networks under 
 Grant. No. FMRX--CT98--0180,  
and by the German Research Council (DFG) under Kr 627/10.


\begin{references}
\bibitem{remark}  In general, the spacing $\Delta_c$ is  random, 
obeying a distribution that is intermediate between the Poisson and
the Wigner-Dyson one. However, in all our results $\Delta_c$
enters under the logarithm, so that the uncertainty in
$\Delta_c$ translates into some numerical factor $\sim 1$
in the argument of the logarithm of a large number.
\end{references}
\end{document}